\documentclass[superscriptaddress, nofootinbib, preprintnumbers, twocolumn]{revtex4}

\usepackage{amsmath}	% AMS Math Package
\usepackage{amsthm}		% Theorem Formatting
\usepackage{amssymb}	% Math symbols such as \mathbb
\usepackage{eufrak}
\usepackage{datetime}
\usepackage{graphicx}
\usepackage{color}
\usepackage{verbatim}
\usepackage{bm}

\usepackage[colorlinks=true, citecolor=midblue, linkcolor=midblue, urlcolor=midblue]{hyperref}

\usepackage{setspace}

\definecolor{grey}{rgb}{0.4,0.4,0.4}
\definecolor{dullmagenta}{rgb}{0.4,0,0.4}
\definecolor{darkblue}{rgb}{0,0,0.4}
\definecolor{midblue}{rgb}{0,0,0.5}
\definecolor{midred}{rgb}{0.5,0,0}
\definecolor{orange}{rgb}{1,0.5,0}
\definecolor{lightbrown}{rgb}{0.75,0.5,0.25}
\definecolor{tan}{cmyk}{0.14,0.42,0.56,0}
\definecolor{djunglegreen}{cmyk}{0.99,0,0.52,0}
\definecolor{lightgreen}{rgb}{0,1,0}
\definecolor{olivegreen}{cmyk}{0.64,0,0.95,0.40}
\definecolor{midgreen}{rgb}{0.0,0.675,0.0}
\definecolor{darkgreen}{rgb}{0,0.5,0}

\newcommand{\vs}{\vspace}

\renewcommand{\.}{\hspace{0.5mm}}

\newcommand{\Drm}{\ensuremath{\mathrm{D}}}

\newcommand{\Frm}{\ensuremath{\mathrm{F}}}
\newcommand{\Grm}{\ensuremath{\mathrm{G}}}
\newcommand{\Hrm}{\ensuremath{\mathrm{H}}}

\newcommand{\Prm}{\ensuremath{\mathrm{P}}}

\newcommand{\Vrm}{\ensuremath{\mathrm{V}}}

\newcommand{\brm}{\ensuremath{\mathrm{b}}}

\newcommand{\erm}{\ensuremath{\mathrm{e}}}

\newcommand{\grm}{\ensuremath{\mathrm{g}}}

\newcommand{\nrm}{\ensuremath{\mathrm{n}}}

\newcommand{\prm}{\ensuremath{\mathrm{p}}}

\newcommand{\Ocal}{\ensuremath{\mathcal{O}}}

\renewcommand{\d}{\ensuremath{\mathrm{d}}}
\newcommand{\eV}{\ensuremath{\, \mathrm{eV}}}

\newcommand{\MeV}{\ensuremath{\, \mathrm{MeV}}}
\newcommand{\GeV}{\ensuremath{\, \mathrm{GeV}}}
\newcommand{\TeV}{\ensuremath{\, \mathrm{TeV}}}

\newcommand{\cm}{\ensuremath{\, \mathrm{cm}}}

\newcommand{\cf}{cf.}

\smallskipamount
\settimeformat{ampmtime}

\usepackage{float}

\begin{document}

%%%%%%%%%%%%%%%%%%%%%%%%%%%%%%%%%%%%%%%%%%%%%%%%%%%%%%%%
\title{Compact Dark Matter Objects via $\bm{N}$ Dark Sectors}

\author{Gia Dvali}
\affiliation{
	Arnold Sommerfeld Center,
	Ludwig-Maximilians-Universit{\"a}t,
	Theresienstra{\ss}e 37,
	80333 M{\"u}nchen,
	Germany,}
\affiliation{
	Max-Planck-Institut f{\"u}r Physik,
	F{\"o}hringer Ring 6,
	80805 M{\"u}nchen,
	Germany}

\author{Emmanouil Koutsangelas}
\email{emi@mpp.mpg.de}
\affiliation{
	Arnold Sommerfeld Center,
	Ludwig-Maximilians-Universit{\"a}t,
	Theresienstra{\ss}e 37,
	80333 M{\"u}nchen,
	Germany,}
\affiliation{
	Max-Planck-Institut f{\"u}r Physik,
	F{\"o}hringer Ring 6,
	80805 M{\"u}nchen,
	Germany}

\author{Florian K{\"u}hnel}
\affiliation{
	The Oskar Klein Centre for Cosmoparticle Physics,\\
	Department of Physics,
	Stockholm University,
	AlbaNova University Center,\\
	Roslagstullsbacken 21, SE-10691 Stockholm, Sweden}

\date{\formatdate{\day}{\month}{\year}, \currenttime}

%%%%%%%%%%%%%%%%%%%%%%%%%%%%%%%%%%%%%%%%%%%%%%%%%%%%%%%%
\begin{abstract}
We propose a novel class of compact dark matter objects in theories where the dark matter consists of multiple sectors. We call these objects $N$-MACHOs. In such theories neither the existence of dark matter species nor their extremely weak coupling to the observable sector represent additional hypotheses but instead are imposed by the solution to the Hierarchy Problem and unitarity. The crucial point is that particles from the same sector have non-trivial interactions but interact only gravitationally otherwise. As a consequence, the pressure that counteracts the gravitational collapse is reduced while the gravitational force remains the same. This results in collapsed structures much lighter and smaller as compared to the ordinary single-sector case. We apply this phenomenon to a dark matter theory that consists of $N$ dilute copies of the Standard Model. The solutions do not rely on an exotic stabilization mechanism, but rather use the same well-understood properties as known stellar structures. This framework also gives rise to new microscopic superheavy structures, for example with mass $10^8$\,g and size $10^{-13}$\,cm. By confronting the resulting objects with observational constraints, we find that, due to a huge suppression factor entering the mass spectrum, these objects evade the strongest constrained region of the parameter space. Finally, we discuss possible formation scenarios of $N$-MACHOs. We argue that, due to the efficient dissipation of energy on small scales, high-density regions such as ultra-compact mini-halos could serve as formation sites of $N$-MACHOs.
\end{abstract}

%%%%%%%%%%%%%%%%%%%%%%%%%%%%%%%%%%%%%%%%%%%%%%%%%%%%%%%%
\maketitle

%%%%%%%%%%%%%%%%%%%%%%%%%%%%%%%%%%%%%%%%%%%%%%%%%%%%%%%%
%%%%%%%%%%%%%%%%%%%%%%%%%%%%%%%%%%%%%%%%%%%%%%%%%%%%%%%%
\section{Introduction}
\label{sec:Introduction}
\vs{-5mm}
To the present day the fundamental nature of the dark matter (DM) remains one of the major mysteries in cosmology and particle physics. At the same time, another major unanswered question in particle physics is what stabilizes the mass term of the Higgs boson against quantum corrections at a value more than 32 orders of magnitude below the Planck mass. This puzzle is known as the {\it Hierarchy Problem}. Not surprisingly, the above two mysteries are the main motivation behind many attempts to extend the Standard Model (SM). Obviously, most interesting are the theories that address both puzzles simultaneously and establish an underlying connection among them. Such scenarios are extremely rare. In most of the models, new long-lived particles are introduced by hand with sole purpose to account for the dark matter without addressing any other problem. Even in low-energy supersymmetry, which solves the Hierarchy Problem, the stability of the dark matter particle is not automatic and requires the postulation of an additional symmetry such as $R$-parity. Of course, we cannot exclude that Nature could provide a dark matter candidate without any additional purpose but as a theoretical guideline we wish to give priority to models that address more than one problem.\footnote{An example of the theory that obeys such a standard is that of the axion \cite{Wilczek:1977pj, Weinberg:1977ma}, which is motivated by the Peccei-Quinn solution \cite{Peccei:1977ur} to strong-CP problem and as a byproduct provides an excellent dark matter candidate.}
 
In the present paper we shall focus on such framework which goes under the name of many-species solution to the Hierarchy Problem \cite{Dvali:2007hz}. The key point is that in any low-energy effective theory that includes $N_{\rm tot}$ particle species, the quantum-gravitational cutoff is given by,
	\begin{equation} \label{eq:M*}
		M_{*}
			\equiv
				\frac{ M_{\Prm} }{ \sqrt{N_{\rm tot}\,}\. }
		\; ,
	\end{equation}
where $M_{\Prm}$ is the Planck mass. This is a fully non-perturbative bound that follows from black-hole physics and its derivation can be found in Refs.~\cite{Dvali:2007hz, Dvali:2007wp, Dvali:2008fd, Dvali:2008ec, Dvali:2010vm}. Hence, if the theory would contain $N_{\rm tot} \sim 10^{32}$ different species, the fundamental gravity scale would be $M_{*} \sim 1\TeV$. This fact nullifies the Hierarchy Problem, since the UV-sensitive radiative corrections to the Higgs mass arising from Standard Model loops must be cutoff at the scale $M_{*}$.

Notice, new species must be extremely weakly interacting with the observable particles from the SM. It is important that this is not an additional assumption but is a constraint coming from unitarity \cite{Dvali:2007wp, Dvali:2008fd}. This fact makes the hidden species into natural dark matter candidates. In a particularly minimalistic scenario the new species represent the exact copies of the Standard Model \cite{Dvali:2007hz, Dvali:2007iv, Dvali:2009ne, Dvali:2009fw}. 

We would like to note that, as explained in \cite{Dvali:2007iv}, the latter scenario provides an additional bonus by addressing the strong-CP problem without need of an axion. The idea is that the sum of CP-violating order parameters over all $10^{32}$ copies of QCD is bounded from above by the cutoff scale. This sets the value of the QCD $\theta$-angle in the observable sector in an experimentally interesting range below the current phenomenological bound. 

Before proceeding, we cite some other implications of this framework studied in Refs.~\cite{Dvali:2008jb, Dvali:2008rm, Rahman:2009rq, Kovalenko:2010ti, Kovalenko:2010qv, Addazi:2016fbj, Cohen:2018cnq, delRio:2018vrj, Han:2018pek, Archer-Smith:2019gzq}. We also note that a possible solution with less number of copies was discussed in Ref. \cite{Arkani_Hamed_2016}. Below, we shall try to keep our discussion maximally general and to distill the model-independent predictions. We shall assume a framework in which there exists a large number of hidden sectors that interact with the Standard Model as well as among each other extremely weakly. At the same time, the interactions within a given sector are assumed to be relatively stronger than among different sectors.

The solution of the Hierarchy Problem demands that the total number of species is $N_{\rm tot} \sim 10^{32}$ \cite{Dvali:2007hz}. However, what matters for their astrophysical consequences is the number of species that become significantly populated during the cosmological evolution. We shall refer to these as {\it actualized species} and denote their number by $N$. Obviously, theoretically $N \leqslant N_{\rm tot}$. However, in practice the number of actualized species could be much smaller.

The reason is the following. As was shown in Ref.~\cite{Dvali:2007wp} and in more details in Ref.~\cite{Dvali:2008fd}, unitarity imposes severe constraints on the strengths of the couplings among different species. The point is that for any particle the maximal possible strength of interaction with the majority of species is suppressed by $M_{*}^{2} / M_{\Prm}^{2} = 1 / N_{\rm tot}$. That is, the particles from a given sector, e.g.~the SM, are allowed to couple relatively strongly only with the few ``privileged neighbors'', while the coupling to the rest is extremely weak (i.e.,~$1 / N_{\rm tot}$-suppressed). In the opposite case, unitarity would be violated by loop expansions and scattering amplitudes.

In other words, if we think of the species label $j\mspace{1mu}=\mspace{1mu}1, \ldots, N_{\rm tot}$ as a discrete coordinate in an imaginary {\it space of species}, the unitarity imposes a certain well-defined sense of locality and distance in this space \cite{Dvali:2008fd}. 

The above knowledge has important implications once the many-species framework is embedded in the cosmological context of the inflationary scenario \cite{Dvali:2009fw} (see also Refs.~\cite{Arkani_Hamed_2016, Cohen:2018cnq, delRio:2018vrj, Han:2018pek, Archer-Smith:2019gzq} for some related analysis and \cite{Watanabe:2007tf} for gravitational inflaton decay).

As it is well-known, in the latter scenario, the key player is a slowly-rolling inflaton field that serves as a clock controlling the end of inflation. Through its oscillations and subsequent decay, this field also ensures the conversion of the vacuum energy density into radiation. Since, as the fact of Nature, our sector is most densely populated, the inflaton field must couple strongest to our sector. After inflation, this field decays and populates the Universe both with ordinary SM particles as well as with the hidden sector species. 

Of course, other sectors could posses their own inflaton candidates. In particular, in the scenario with exact copies of the SM, this is imposed by the permutation symmetry. In this case, a model builder has to decide whether the inflaton field is a singlet under the permutation group or transforms under it non-trivially. 

The former possibility would imply that the inflaton couples to all the copies with the universal strength which by unitarity must be suppressed by $1/N_{\rm tot}$. Consequently, in such a scenario after reheating all the sectors would end up being equally densely populated, which is excluded observationally. The alternative possibility is that inflaton transforms under the permutation symmetry and thus shares the label $j$ with the SM copies. That is, each SM copy possesses its own inflaton candidate. However, the permutation symmetry is spontaneously broken by the cosmological evolution and the sector that drives the latest stage of inflation is the one that matters for the reheating. By default, the sector that this particular inflaton populates most efficiently is the one that we today call ``our" SM copy. Of course, equally legitimate are the scenarios in which the hidden sectors are not related by any permutation symmetry with our sector and are very different. All the above-mentioned unitarity constraints, however, hold even in such cases and the reheating process must obey them.

The question now is: how densely are the other sectors populated? The relevant decay and scattering processes are: 
\begin{itemize}

	\item Inflaton $\rightarrow$ Us,
	
	\item Inflaton $\rightarrow$ Them,
	
	\item Us $\rightarrow$ Them,
	
\end{itemize} 
where obviously ``us" and ``them" denote the SM and hidden sectors, respectively. Notice, the process ``them$\,\rightarrow\,$us'' can be dismissed due to very strong dilution of the individual hidden sectors.
Thus, there exist two mechanisms for populating the dark sectors: 
\begin{enumerate}

	\item {\it Direct}: Through the decay of the inflaton into the\\
		\phantom{\it Direct: }dark species (Inflaton $\rightarrow$ Them).

	\item {\it Indirect}: Through re-scattering of particles of the\\
		\phantom{\it Indirect: }visible sector into the dark ones\\
		\phantom{\it Indirect: }(Inflaton $\rightarrow$ Us $\rightarrow$ Them).

\end{enumerate}

Notice, although the latter mechanism of populating dark sector can also be operative in standard single-sector dark matter scenarios, it is the many-species framework that makes it especially efficient due to enormous $1 / N_{\rm tot}$-suppression of the inverse process.

The direct population process is determined by the decay rate $\Gamma_{{\rm infl}\,\rightarrow\,{\rm them} } = \sum_{j\mspace{1mu}=\mspace{1mu}1}^{N_{\rm tot}} \Gamma_{{\rm infl}\,\rightarrow\,j}$, where $\Gamma_{{\rm infl}\,\rightarrow\,j}$ represents the decay rate of the inflaton into the $j$-th hidden species. Likewise, the indirect process ``us $\rightarrow$ them'' is controlled by the total re-scattering of SM quanta into the hidden sectors that is given by summing over the individual scattering rates $\Gamma_{{\rm us}\,\rightarrow\,{\rm them} } = \sum_{j\mspace{1mu}=\mspace{1mu}1}^{N_{\rm tot}} \Gamma_{{\rm us}\,\rightarrow\,j}$. The unitarity puts certain generic constraints such as $\Gamma_{\rm infl} \equiv \Gamma_{{\rm infl}\,\rightarrow\,{\rm them} } + \Gamma_{{\rm infl}\,\rightarrow\,{\rm us} }\,\lesssim m_{\rm infl}$, where $m_{\rm infl}$ is the mass of the inflaton. This constraint means that during the reheating process the inflaton represents a weakly-interacting degree of freedom and can be consistently treated within effective field theory without the risk of violating perturbative unitarity. 

The second constraint is $\Gamma_{{\rm us}\,\rightarrow\,{\rm them}} \lesssim T_{\rm SM}$, where $T_{\rm SM}$ is the temperature in the SM sector at which the re-scattering into hidden species takes place. Furthermore, in order not to destroy the success of standard big bang nucleosynthesis, we assume that the reheating temperature $T_{\rm R}$ is higher than the big bang nucleosynthesis temperature, $\sim$\MeV. Then, at the moment of reheating, the rates must satisfy: $\Gamma_{{\rm infl}\,\rightarrow\,{\rm them} } \ll {\Gamma_{{\rm infl}\,\rightarrow\,{\rm us} }}$. This is because we know that during nucleosynthesis Universe's energy balance was dominated by the energy density of radiation of the SM species. Correspondingly, the total energy density deposited by the inflaton into the hidden sector must be negligible as compared to the energy density deposited into the SM particles. Of course, later, after the Universe cools down and crosses into matter domination, the integrated energy density of matter from the hidden sector must become a few times larger than the density in the observable sector, in order to account for current dark matter density.

Subject to the above constraints we can consider different cosmological scenarios. One regime is achieved when only a relatively small number of sectors $N \ll N_{\rm tot}$ is actualized while the rest remains empty. Another possibility is to assume that no privileged neighbours exist and that all the dark sectors are equally populated. In the latter case, each dark sector is $\sim 10^{-32}$ times more dilute than our sector. Of course, more involved scenarios in which the partial density is some non-trivial function of species index $j$ are also possible. 
 
In the present paper, we shall try to keep our discussion maximally open, while keeping in mind the above general constraints. For a more detailed discussion of concrete cosmological realizations of many-species dark matter, we refer the reader to Ref.~\cite{Dvali:2009fw}.

The crucial difference between these theories and traditional DM models is that DM is able to dissipate energy due to the intrinsic interactions in each sector, possibly enabling the formation of non-baryonic compact objects. Analogous possibilities were already realized in earlier papers \cite{ArkaniHamed:1999zg, Dvali:2009ne} without any detailed analysis.

In the current paper we shall investigate how various dark sectors can \textit{collectively} form compact objects. We shall call these objects $N$-MACHOs. Furthermore, we will remain agnostic about the origin of the different dark sectors. We simply consider $N$ gravitationally-coupled dark sectors with intrinsic SM-strength interactions.

Intuitively, it is clear that the mass spectrum of $N$-MACHOs must be different from that of a single-sector object. Putting the particles from the $N$ sectors into a certain volume, gravity will try to collapse the system while the pressure from each sector will try to counteract the collapse. Eventually, the total pressure can balance gravity, leading to an equilibrium configuration. The fact that in this case each particle can only interact with a small fraction of the present particles, leads to a strong reduction of the pressure as compared to the single sector case. At the same time, gravitational forces remain unaffected. In other words, each particle feels the pressure only from the species of its own sector, while it experiences a universal gravitational attraction from all the other sectors. Consequently, an equilibrium configuration can only be achieved for much smaller masses and thus much smaller radii than in ordinary single-sector models.

Compact DM structure not only represent a powerful tool to discriminate between classes of DM models, but could also provide new information on the spectrum of inflation. For instance, this is the case for primordial black holes (PBHs) (see Ref.~\cite{Carr:2016drx} for a review). The standard scale-invariant inflationary power spectrum results in essentially zero PBHs in the present day \cite{PhysRevD.50.7173}. Therefore, the discovery of such an object would require an additional feature in the standard power spectrum. $N$-MACHOs behave similar but are easier to produce than PBHs since they do not require that high primordial overdensities.

Moreover, in light of recent discoveries by the Laser Interferometer Gravitational Wave Observatory (LIGO) \cite{Abbott:2016blz, Abbott:2016nmj}, a promising signature resides in the form of gravitational waves. If future generations of gravitational-wave experiments would observe a signal from a compact object merger with neutron-star-like compactness but different mass, then a new kind of compact object must exist. This object would most likely consist of non-baryonic DM, due to the lack of alternatives in the SM. Additionally, if the number of violent astrophysical processes happening today in the dark sector is of the same order as those happening in ours, then the total number of gravitational-wave sources increases by the ratio of DM to ordinary matter \cite{ArkaniHamed:1999zg}.

This paper is structured as follows: In Sec.~\ref{sec:N--MACHOs} the Einstein equations are solved, resulting in the solution we call $N$-MACHOs. The calculations are first performed for equal particle content and equal particle density in each sector. Later they are generalized to non-equal particle content and particle density. Section \ref{sec:N--Protonium} discusses some novel stable bound-states in the particle spectrum that emerge in the many species scenario. In Sec.~\ref{sec:Radiation-into-Dark-Photons} we elaborate on the possibility of radiation into dark photons during merger processes involving $N$-MACHOs. In Sec.~\ref{sec:Constraints-and-Signatures} the constraints on $N$-MACHOs are presented, while Sec.~\ref{sec:Formation} discusses potential formation scenarios. Finally, in Sec.~\ref{sec:Conclusion-and-Outlook} we summarize our results and give an outlook.

%%%%%%%%%%%%%%%%%%%%%%%%%%%%%%%%%%%%%%%%%%%%%%%%%%%%%%%%
%%%%%%%%%%%%%%%%%%%%%%%%%%%%%%%%%%%%%%%%%%%%%%%%%%%%%%%%
\section{$N$-MACHOs}
\label{sec:N--MACHOs}

%%%%%%%%%%%%%%%%%%%%%%%%%%%%%%%%%%%%%%%%%%%%%%%%%%%%%%%%
\subsection{Hydrostatic Equilibrium}
\label{sec:Hydrostatic-Equilibrium-in-General-Relativity}

In this Subsection, we discuss the equations of hydrostatic equilibrium in General Relativity in the presence of $N$ matter sectors.

First, it should be noted that the Einstein field equations only depend on the total energy-momentum tensor. By assumption, the different sectors interact only via gravity, which at the energies of interest is extremely weak
	\footnote{
		This is true even though the gravitational self coupling and/or coupling among few nearest neighbour species become strong around energies $M_* \sim$ TeV. The reason is the same as the one explained in \cite{ArkaniHamed:1998rs} in the context of large extra dimensions: Strengthening gravity at distance of $M_*^{-1} \sim 10^{-17}$cm has a negligible effect on the gravitational dynamics of a macroscopic object of size $R \gg M_*^{-1}$. For such objects, the treatment of gravity within standard Newton-Einstein theory is a justified approximation.}. 
Therefore, with a good approximation, the total energy-momentum tensor can be decomposed as,
	\begin{equation}
		( T_{\rm tot} )^{}_{\mu\nu}
			=
				\sum_{j\mspace{1mu}=\mspace{1mu}1}^{N} T^{}_{\mu\nu, j}
				\; ,
	\end{equation}
where the index $j$ labels the individual sectors (or species). Following the standard approach, we will assume that each sector behaves as a perfect fluid, which in the rest frame
takes the following form,
	\begin{align}
	\begin{split}
		( T_{\rm tot} )_{\mu\nu}
			&=
				\text{diag}\mspace{-1mu}
				\left(
					-\rho_{\rm tot},
					P_{\rm tot},
					P_{\rm tot},
					P_{\rm tot}
				\right)
				\\[1mm]
			&=
				\sum_{j\mspace{1mu}=\mspace{1mu}1}^{N}\text{diag}\mspace{-1mu}
				\left(
					- \rho_{j},\,
					P_{j},\,
					P_{j},\,
					P_{j}
				\right)
			\, ,
	\end{split}
	\end{align}
where $\rho_{\rm tot}$ and $P_{\rm tot}$ are the total energy density and the total pressure, respectively.

Solving Einstein's equations with the standard spherical ansatz (i.e., neglecting rotation) yields the Tolman-Oppenheimer-Volkoff (TOV) equation
	\begin{align}\nonumber
		\frac{\d P_{\rm tot}( r )}{\d r}
			=
				&-
				\frac{M( r )}{M_{\Prm}^{2}\.r^{2}}\;
				\rho_{\rm tot}\mspace{1mu}
				\bigg(
					1 + \frac{P_{\rm tot}( r )}{\rho_{\rm tot}( r )}
				\bigg)
				\\[1.5mm]
				&\times\bigg(
					1 + \frac{4\pi\.r^{3}\.P_{\rm tot}( r )}{M( r )}
				\bigg)\!
				\bigg(
					1 - \frac{2 M( r )}{M_{\Prm}^{2}\.r}
				\bigg)^{\!-1}
				\; ,
	\end{align}
where
	\begin{equation}
		M( r )
			\equiv
				\int_{0}^{r} \d r'\;
				4\pi\.r'^{2} \rho_{\rm tot}(r')
				\; .
	\end{equation}
The TOV equation is the fundamental equation of Newtonian astrophysics with General Relativity corrections supplied by the three brackets. In order to solve it, an equation of state (EOS), i.e., the pressure $P_{\rm tot}( S_{\rm tot}, \rho_{\rm tot} )$ as a function of the entropy $S_{\rm tot}$ and the density $\rho_{\rm tot}$, needs to be specified.

In this article, we will only consider polytropes in the Newtonian limit -- so called Newtonian polytropes -- since these provide a simplified model for the stellar interior. They are characterized by a polytropic EOS, i.e.,
	\begin{equation} \label{eq:PolytropicEOS}
		P_{\rm tot}	
			=
				K^{}_{\rm tot}\;
				\rho_{\rm tot}^{\gamma}
				\; ,
	\end{equation}
where $\gamma$ and $K_{\rm tot}$ are constants. Physically speaking, a polytropic EOS describes an object with uniform entropy, which, for example, is the case if the latter has effectively zero temperature or is in convective equilibrium. These properties may sound highly idealistic, but they often give a right order-of-magnitude estimate. For example, the former condition is given in white dwarfs or neutron stars, while the latter is realized in supermassive stars. Furthermore, we assume that the chemical composition is uniform, so that one polytropic EOS describes the whole object.

All properties of the $N$-MACHOs are then determined by the central density $\rho( 0 )$ and the only free parameter in our theory, $N$. The non-relativistic TOV equation in the presence of $N$ sectors has exactly the same overall form as in case of a single sector. Therefore, the $N$-sector solution will be of the same form as the single-sector one. In other words, the $N$-sector solution is also a Newtonian polytrope. Since these are well-studied astrophysical objects, we summarize respective details in App.~\ref{sec:Newtonian-Polytropes} (see for instance Ref.~\cite{Weinberg1972-WEIGAC} for a pedagogical treatment).

%%%%%%%%%%%%%%%%%%%%%%%%%%%%%%%%%%%%%%%%%%%%%%%%%%%%%%%%
\subsection{Solution for Equal Sectors}
\label{sec:Solution-for-Identical-Sectors}

As stated previously, a polytropic EOS can be thought of as describing an isentropic stellar object. The key point now is, that if an object composed of $N$ exclusively gravitationally-interacting components is isentropic, then each component has to be isentropic as well. Consequently, each sector and the full system possess a polytropic EOS with the same exponent $\gamma$. Let us, for now, assume an equal particle content in each sector, or equivalently, the universal polytropic constant $K$. That is,
	\begin{equation}
		P_{j}
			=
				K
				\rho_{j}^{\gamma}
				\; .
	\end{equation}
In the presence of $N$ purely gravitationally-interacting sectors, the total pressure satisfies,
	\begin{equation}
		P_{\rm tot}
			=
				\sum_{j\mspace{1mu}=\mspace{1mu}1}^{N}\.P_{j}
			=
				K \.
				\sum_{j\mspace{1mu}=\mspace{1mu}1}^{N}\.
				\rho_{j}^{\gamma}
			\equiv
				K^{}_{\rm tot}\.\rho_{\rm tot}^{\gamma}
				\; .
	\end{equation}
From this we see that the influence of the $N$ sectors is encoded in the quantity
	\begin{equation} \label{eq:NInfluence}
		K_{\rm tot}
			=
				\frac{K\. \sum_{j\mspace{1mu}=\mspace{1mu}1}^{N}\rho_{j}^{\gamma} }{ \rho_{\rm tot}^{\gamma} }
				\; .
	\end{equation}
Assuming that the sectors have equal particle densities, i.e.,
	\begin{equation} \label{eq:rhoEPS}
		\rho_{\rm tot}
			=
				\sum_{j\mspace{1mu}=\mspace{1mu}1}^{N}\.
				\rho_{j}
			\equiv
				N \rho
				\; ,
	\end{equation}
we have
	\begin{equation} \label{eq:ESFactor}
		K_{\rm tot}
			=
				\frac{K\.\.}{N^{\gamma - 1 }}
		\; .
	\end{equation}
Inserting this in the expressions for the radius and the mass of a generic Newtonian polytropic object (see Eqs.~\eqref{eq:PolR} and \eqref{eq:PolM}, respectively in App.~\ref{sec:Newtonian-Polytropes}) we finally arrive at
	\begin{subequations}
	\begin{align}
		R
			&\sim
				R_{0}\.
				\frac{ 1 }{ \sqrt{N\.}\. }
			\;,
	\label{eq:REPS}
			\displaybreak[1]
			\\[1.5mm]
	\intertext{and}
		M
			&\sim
				M_{0}\.
				\frac{ 1 }{ \sqrt{N\.}\. }
				\; ,
	\label{eq:MEPS}
	\end{align}
	\end{subequations}
where $R_{0}$ and $M_{0}$ depend on the particle content and the nature of pressure in each sector. Thus,
increasing the number of sectors reduces the mass of a stellar object by a factor $1 / \sqrt{N\.}$. What is remarkable is that this factor is independent of $\gamma$ and thus is universal for all possible stellar dark matter structures. Given the available range for $N$ motivated by the Hierarchy Problem \cite{Dvali:2009fw}, the potential impact is dramatic.

Let us now apply the above results to the case of $N$ exact SM copies. In this theory the interactions among the particles of a given copy are identical to those of the SM. At the same time, the particles from different copies talk to each other only through gravity.

In this framework the simplest example of $N$-MACHO is provided by the polytropic model of an ``$N$-neutron star". The pressure is dominated by the Fermi pressure $P_{\Frm}$ of $N$ identical sets of non-relativistic copies of neutrons. The masses of hidden neutrons are of course equal to the mass $m_{\nrm}$ of the ordinary SM neutron. Thus, each neutron is kept in the bound-state due to the gravitational attraction from the rest. However, it is balanced against the collapse due to the Fermi pressure exclusively from the fellow neutrons belonging to the same SM copy. At the same time, due to the effectively vanishing temperature of the degenerate neutron gas, the EOS is isentropic and therefore $P_{\Frm}$ is only a function of $\rho$.

In the non-relativistic limit the Fermi momentum $p_{\Frm} = (3\pi^{2} \rho / m_\nrm)^{1/3}$ satisfies $p_{\Frm} \ll m_{\nrm}$ or equivalently $\rho \ll \rho_{\rm cr}$, where $\rho_{\rm cr}$ is the critical density at which $p_{\Frm}=m_{\nrm}$, and the EOS is given by
	\begin{equation}
		P_{\Frm}
			=
				K_{\Frm}\,\rho^{ 5 / 3 }
				\; ,
		\qquad
		K_{\Frm}
			\equiv
				\frac{ 1 }{ 15 \pi^{2}\.m_{\nrm} }\!
				\left(
					\frac{ 3\pi^{2} }{ m_{\nrm} }
				\right)^{\!5/3}
				\; .
	\label{eq:Kdef}
	\end{equation}
This is the EOS of a Newtonian polytrope with polytropic constant $K_{\Frm}$ and polytropic index $\gamma = 5/3$. However, since the quantities appearing in the non-relativistic TOV equation are the total pressure and the total energy density, the total EOS of the system is needed. Summing over all the copies gives
	\begin{equation}
		P_{\rm tot}
			=
				N\.P_{\Frm}
			=
				N\.K_{\Frm}\.
				\rho^{5 / 3}
			=
				\frac{ 1 }{ N^{2/3} }\.K_{\Frm}\.
				\rho_{\rm tot}^{5 / 3}
				\; ,
	\end{equation}
which, as expected, again yields a Newtonian polytrope. For the latter (using again App.~\ref{sec:Newtonian-Polytropes}) we obtain
	\begin{subequations}
	\begin{equation} \label{eq:RofN}
		R
			\sim
				10^{6}\cm\mspace{-2mu}
				\left(
					\frac{ \rho( 0 ) }{ \rho_{\rm cr} }
				\right)^{\mspace{-5mu}-1 / 6}
				\frac{ 1 }{ \sqrt{N\.} }
				\; ,
	\end{equation}
and
	\begin{equation} \label{eq:MofN}
		M	
			\sim
				\.M_{\odot}\mspace{-2mu}
				\left(
					\frac{ \rho( 0 ) }{ \rho_{\rm cr} }
				\right)^{\mspace{-5mu}1 / 2}
				\frac{ 1 }{ \sqrt{N\.} }
				\; .
	\end{equation}
	\end{subequations}
With this, the total density is given by
	\begin{equation}
		\rho_{\rm tot}
			=
				\frac{ M }{ \frac{ 4 \pi }{ 3 } R^{3} }
			\sim
				10^{15}\,
				\frac{ \grm\.\. }{ \cm^{3} }\mspace{-2mu}
				\left(
					\frac{ \rho( 0 ) }{ \rho_{\rm cr} }
				\right)
				N
			\; ,
	\end{equation}
while the density per sector is
	\begin{equation}
		\rho
			=
				\frac{ \rho_{\rm tot} }{ N }
			=
				10^{15}\,\frac{ \grm\.\. }{ \cm^{3} }\mspace{-2mu}
				\left(
					\frac{ \rho( 0 ) }{ \rho_{\rm cr} }
				\right)
			\. .
	\end{equation}
The central density of each sector should be around nuclear density, $\rho_{\rm nucl} \approx 10^{14}\,{\grm} / \cm^{3}$, in order for the degeneracy pressure to be dominant. This roughly sets
	\begin{equation}
		\frac{ \rho( 0 ) }{ \rho_{\rm cr} }
			\lesssim
				10^{-1}
			\. .
	\end{equation}
As in the single-sector case, which would reproduce an ordinary neutron star, the aforementioned total density is close to that of a black hole. For $N = 1$, the usual neutron stars are retrieved, while for $N\,\rightarrow\,\infty$ the mass and radius vanish, showing that these objects would not form. For ``$N$-neutron stars", the mass spectrum depends only on the parameter $N$. For example, for the cosmological scenario proposed in Ref.~\cite{Dvali:2009fw} modified with non-annihilating electrons, the window for $N$ is $10^{16} < N < 10^{32}$. Correspondingly, the ranges for the mass and the size are respectively given by
\vs{1mm}
\begin{subequations}
	\begin{align}
		10^{17}\,\text{\grm} < M < 10^{25}\,{\grm}
			\; ,
			\\[2mm]
		10^{-10}\cm < R < 10^{-2}\cm
			\; .
	\end{align}	
\end{subequations}
We would like to stress that other cosmological scenarios would lead to different windows.

%%%%%%%%%%%%%%%%%%%%%%%%%%%%%%%%%%%%%%%%%%%%%%%%%%%%%%%%
\subsection{$N$-MACHO Solution for Unequal Sectors}
\label{sec:N--MACHO-Solution-for-Unequal-Sectors}

In this Section we relax the assumption of universality of parameters for different species. Therefore, we assign to a sector $j$ an energy density $\rho_{j}$ and a polytropic constant $K_{j}$. In order to find a generalization of Eq. \eqref{eq:ESFactor} for such a case, we need to assume a particular $j$-dependence of these parameters. That is, we need to assume a particular profile in species space.

Consider first the case with $K_{j} \equiv K$, but $\rho_{j} \neq \rho$. Then, it is useful to introduce the following parameterization, 	
	\begin{equation}
		\rho_{j}	
			=
				\rho_{\rm max}\,f_{j}
			\; ,
	\end{equation}
where $\rho_\text{max}$ is a maximal density and $f_{j}$ is some positive discrete function of $j$ with $f_{j\mspace{1mu}=\mspace{1mu}j_{\rm max}}= 1$ and $f_{j\mspace{1mu}\neq\mspace{1mu}j_{\rm max}} \leq 1$. Next, introducing the following notation,
	\begin{equation}
		\rho_{\rm tot}
			=
				\rho_{\rm max}
				\sum_{j\mspace{1mu}=\mspace{1mu}1}^{N} f_{j}
			\equiv
				\rho_{\rm max}\.N_{\rm eff}
		\; ,
	\end{equation}
we can rewrite the r.h.s of Eq.~\eqref{eq:NInfluence} as
	\begin{align}\nonumber
		\frac{ K\. \sum_{j\mspace{1mu}=\mspace{1mu}1}^{N}\rho_{j}^{\gamma} }
		{ \rho_{\rm tot}^{\gamma} }
			&=
				K\.\rho_{\rm max}^{\gamma}
				\frac{\sum_{j\mspace{1mu}=\mspace{1mu}1}^{N}\.f_{j}^{\gamma} }
				{ \rho_{\rm tot}^{\gamma} }
			\le
				K\.\rho_{\rm max}^{\gamma}
				\frac{\sum_{j\mspace{1mu}=\mspace{1mu}1}^{N}\.f^{}_{j} }
				{ \rho_{\rm tot}^{\gamma} }
				\displaybreak[1]
				\\[1.5mm]
			&=
				K\.\left(
					\frac{\rho_{\rm max}}{\rho_{\rm tot}}
				\right)^{ \mspace{-5mu}\gamma - 1 }
			=
				\frac{ K\.\.}{ N_{\rm eff}^{ \gamma - 1 } }
			\; .
	\end{align}
Since ${N_{\rm eff} \leq N}$, the total pressure can be larger for non-universal sectors. Consequently, by replacing $N$ with $N_{\rm eff}$ in Eqs.~(\ref{eq:RofN},b), it becomes clear that the $N$-MACHOs are heavier and larger in this scenario.

Consider now the case with $K_{j} \neq K$ and $\rho_{j} \neq \rho$. This corresponds to the case in which particles of different sectors have different masses. The total pressure now becomes
	\begin{equation} \label{eq:Ninfl2}
		P_{\rm tot}
			=
				\frac{ \sum_{j\mspace{1mu}=\mspace{1mu}1}^{N}\.K^{}_{j}\.\rho_{j}^{\gamma} }
				{ \rho_{\rm tot}^{\gamma} }\,
				\rho_{\rm tot}^{\gamma}
			\; .
	\end{equation}
Repeating the previous approach, we find
	\begin{align} \label{eq:n= 1}
		\frac{ \sum_{ j\mspace{1mu}=\mspace{1mu}1}^{N}\.K^{}_{j}\.\rho_{j}^{\gamma} }
		{ \rho_{\rm tot}^{\gamma} }
			\le
				\frac{ K_{\rm max} }{ N_{\rm eff}^{\gamma - 1} }
			\; ,
	\end{align}
where $K_{\rm max}$ is defined by $K_{\rm max} \geq K_{j}$ for all $j$. This case also results in heavier and larger objects.

Finally, we must note that even in non-universal case, the most interesting regimes are the ones in which the distribution in $j$ is more or less uniform over the actualized $N$ species. In the opposite case, i.e., when a single sector dominates within a given stellar object, the situation effectively reduces to $N_{\rm eff} \simeq N \simeq 1$ case. This is trivially exhibited by the above calculation. Thus, for instance, the physics of ordinary stellar objects that are dominated by SM baryons is unchanged.

%%%%%%%%%%%%%%%%%%%%%%%%%%%%%%%%%%%%%%%%%%%%%%%%%%%%%%%%
%%%%%%%%%%%%%%%%%%%%%%%%%%%%%%%%%%%%%%%%%%%%%%%%%%%%%%%%
\section{$N$-Protonium}
\label{sec:N--Protonium}

Not surprisingly, having a large number of hidden sectors gives rise to a plethora of new exotic stable bound-states in the particle spectrum. This is because various stable particles from different sectors can become gravitationally bound to each other. These structures can be viewed as the extreme microscopic versions of $N$-MACHOs. As an example, let us consider a bound-state formed out of $N \sim 10^{32}$ species of stable baryons, one from each sector. Namely, in this bound-state each Standard-Model-like sector is represented either by a proton or an anti-proton. Each baryon ``sees" the rest of particles exclusively through gravity and is confined by their collective gravitational potential.

Such a state, which can be called $N$-protonium, has the following characteristics. Its mass is given by $M \sim 10^{32}\,m_{\prm} \sim 10^{32} \GeV \sim 10^{8}\,\grm$, whereas its radius is given by the radius of a proton, $R \sim 10^{-13}\cm$. Although microscopic, this length exceeds the corresponding gravitational Schwarzschild radius $R_{\Grm} \sim 10^{-20}\cm$ by seven orders of magnitude. Correspondingly, the $N$-protonium is stable with respect to collapse into a black hole. The gravitational binding energy experienced by each proton is approximately $V \sim m_{\prm}\.R_{\Grm} / R \sim 10^{2}\eV$. Notice, this is close to an attractive gravitational potential experienced by a proton near the surface of the earth and is much less than the nuclear binding energy. Of course, such binding energy is not sufficient for stabilizing a neutron against the beta-decay. Because of this reason, the stable configuration involves a charged baryon (a proton or an antiproton) from each sector, rather than a neutron.

Notice that because of the microscopic size of the object, the gravitation binding force near the surface of $N$-protonium is 24 orders of magnitude stronger than the analogous gravitational force at the surface of the earth. During a passage through an ordinary macroscopic stellar object, the $N$-baryonium will remain essentially undisrupted. This is because the tidal forces that could potentially destroy the bound-state are much weaker that the attractive force that keeps the baryon species together. For example, the tidal acceleration caused by solar gravity near the surface of the sun is 47 orders of magnitude weaker than the binding acceleration, whereas the tidal acceleration caused by a neutron star is 32 orders of magnitude weaker.

Therefore, the only likely loss that $N$-baryonium can experience during the pass of a stellar object is that it will be stripped of ``our" ordinary proton due to a non-gravitational interaction with the stellar medium. The probability of loss of hidden baryons, due to non-gravitational encounter with their own species, is negligible because of extreme dilution of the hidden sectors.

Now, taking $N$-baryonium as a starting object, we can scan the landscape of states that can be obtained by some simple variations of it. For instance, places of protons from some of the species may be ``vacant", which would result in a lighter bound-state with less number of constituents, i.e., $N < 10^{32}$. Correspondingly, the binding potential for each constituent proton, $V \sim N \, 10^{-30}\eV$, becomes more shallow for smaller $N$. Also, some of the baryons may be substituted by the electrons or positrons from the same sectors. Obviously, it is possible to trivially obtain much more complex structures by putting together more composites from different sectors, such as, e.g., the neutral atoms or the entire chemical elements. We shall not discuss them further. 

In summary, the theory with a large number of dark sectors opens a possibility for the existence of new types of simple and stable composite structures that are microscopic in size but macroscopic in mass. In contrast, in theories with a single non-interacting dark matter species, such composite structures are hard to stabilize against a decay or a collapse.

%%%%%%%%%%%%%%%%%%%%%%%%%%%%%%%%%%%%%%%%%%%%%%%%%%%%%%%%
%%%%%%%%%%%%%%%%%%%%%%%%%%%%%%%%%%%%%%%%%%%%%%%%%%%%%%%%
\section{Radiation into Dark Photons}
\label{sec:Radiation-into-Dark-Photons}

The large number of dilute dark sectors opens up the possibility of enhanced radiation into hidden massless species in various structure-formation processes with the participation of dark matter. An example of such a process would be provided by a merger of binaries with the participation of $N$-MACHOs. The latter objects can pair-up and form binaries either with ordinary stellar objects, such as neutron stars or black holes, or with other $N$-MACHOs. As a result of such merger, a certain fraction of radiated energy will appear unaccounted by ordinary electromagnetic radiation and gravitational waves. For concreteness, we shall illustrate this phenomenon for the scenario with $N$ exact Standard Model copies.

Because of the extreme dilution of particles in these copies, the $N$-MACHOs can easily have an excess of charge with respect to some or all dark electromagnetic gauge symmetries. Existence of charges with respect to the hidden photons, opens up invisible channels of radiation into the corresponding dark sectors. Because of this, the overall rate of energy loss into the dark matter is enhanced by $N$.

This situation is very different from the ordinary stellar structures in two respects. First, in such objects the large fluctuations of the ordinary electric charge are suppressed due to its swift neutralization by readily-available freely-floating charges. Secondly, there is no enhancement by $N$. In the multiple-copy scenario, on the contrary, the fluctuations of the individual charges -- since they source different photon species -- instead of averaging out to zero, add up in the emission rate and sharply enhance it.

In order to perform a simple qualitative estimate, consider a merging binary system with the participation of an $N$-MACHO. We shall be interested in the radiation emitted during the latest stages of merger. Let the characteristic mass of the system be $M$ and its size be $R$. We shall assume that dark matter is predominantly represented by dark protons (anti-protons). While electrons (positrons) do exist in equal numbers, the encounter is unlikely due to an extreme dilution. We also assume that the net charge in each sector (i.e., the number difference between the protons and anti-protons) is significantly larger than one and that the separation amongst the charges exceeds the size of the proton.

Since each proton is attracted by the total mass of the system, the maximal net charge in each copy is given by
	\begin{equation} \label{eq:Q}
 		Q
 			\sim
 				\alpha^{-1}\.R_{\Grm}\.m_{\prm}
 			\; ,
	\end{equation}
where $\alpha \sim 10^{-2}$ is the electromagnetic fine-structure constant of the respective SM copy and is universal for all the sectors. The maximal power of electromagnetic radiation per copy is then given by
	\begin{equation} \label{eq:rate}
		\frac{ \d E }{ \d t }
 			\sim
 				\alpha\.Q^{2}\.\frac{ R_{\Grm}^{2} }{ R^{4} }
 			 			\; .
	\end{equation}
Inserting Eq.~\eqref{eq:Q} and multiplying by $N$, we obtain the total power of dark electromagnetic radiation in all the copies:
\begin{equation} \label{eq:PD}
	P^{}_{\Drm}
 			\sim
 				N\,\frac{ m_{\prm}^{2} }{ \alpha }\,
				\frac{ R_{\Grm}^{4} }{ R^{4} }
 			\; .
	\end{equation}
Now, taking into account that the power of ordinary gravitational radiation is given by
	\begin{equation} \label{eq:Pgr}
 		P_{\rm gr} \equiv \frac{ \d E }{ \d t }\Big|_{\rm gravity}
 			\sim
 				M\,\frac{ R_{\Grm}^{4} }{ R^{5} }
 			\; ,
	\end{equation}
we arrive to the following fraction
	\begin{equation} \label{eq:epsilonD}
 		\epsilon^{}_{\Drm}
			\equiv
				\frac{ P_{\Drm} }{ P_{\rm gr} }
 			\sim
 				\frac{ m_{\prm}^{2}\.R }{ \alpha\.M }\,N
 			\; .
	\end{equation}
Defining the gravitational analog of the fine-structure constant for the proton, $\alpha_{\rm gr} \equiv m_{\prm}^{2} / M_{\rm Pl}^{2}$, we can rewrite the above expression in a much more transparent and universal form,
\vs{-1mm}
	\begin{equation} \label{eq:epsilonD2}
 		\epsilon^{}_{\Drm}
 			\sim
 				\frac{ \alpha_{\rm gr} }{ \alpha }\,
				\frac{ R }{ R_{\Grm} }\,N
 			\; .
	\end{equation}
This expression applies for charge carrying dark matter species of arbitrary mass $m$, provided $\alpha_{\rm gr}$ is understood as the gravitational coupling. If the net charge is due to dark protons, $m_{\prm} \sim \GeV$, we have $\alpha_{\rm gr} \sim 10^{-37}$ and correspondingly the maximal fraction of dark electromagnetic radiation is given by,
	\begin{equation} \label{eq:epsilonDgeneral}
 		\epsilon^{}_{\Drm}
 			\sim
 				10^{-37}\,\frac{ R }{ R_{\Grm} }\.N
 			\; .
	\end{equation}
For an $N$-MACHO with $R$ close to $R_\Grm$ and $N \sim 10^{32}$ this gives $\epsilon^{}_{\Drm} \sim 10^{-4}$. For less compact objects this would of course be higher. Notice that the estimate (\ref{eq:Q}) does not apply to objects in which the charge is stabilized by other forces. For example, the size of $N$-protonium is set by the size of the proton, $R \sim 10^{-13}\cm$, which is determined by the strong interaction scale. Such $N$-protonium gravitationally merging with a neutral object of approximately the same mass and size would radiate energy $12$ orders of magnitude larger into dark photons than in gravitons.

As an alternative example, consider the case when dark matter is represented by $N \sim 10^{32}$ species of $m \sim 1 \TeV$ mass, each carrying a unit charge under the respective $U(1)$ gauge symmetry, with gauge coupling $\alpha_D$. Now, consider an $N$-protonium-type object but instead composed out of these TeV-mass species, one from each $10^{32}$ sectors placed on top of each other within the size of the Compton wavelength $R \sim m^{-1}$. Such object will have a mass $M \sim 10^{32}\TeV\sim 10^{11}$\,g and unit charges (of either sign) with respect to dark $U(1)$s. Notice that since $m$ is very close to cutoff, the gravitational radius of such an object is close to its radius, $ R \sim R_{\Grm} \sim m^{-1} \sim 10^{-17}\cm$. During the merger or collision of two such objects we get $\epsilon_D \sim \alpha_D$. 

Of course, one can play around with many analogous structures but the general point we would like to make is that the multiplicity of charged dark matter species allows to increase the intensity of dark radiation at the expense of opening up new channels that are not in conflict with gravitational binding force.

%%%%%%%%%%%%%%%%%%%%%%%%%%%%%%%%%%%%%%%%%%%%%%%%%%%%%%%%
%%%%%%%%%%%%%%%%%%%%%%%%%%%%%%%%%%%%%%%%%%%%%%%%%%%%%%%%
\section{Constraints \& Signatures}
\label{sec:Constraints-and-Signatures}

The question of how to pose constraints, or detection prospects, on $N$-MACHOs overlaps with that for general compact objects at similar masses, such as primordial black holes (PBHs) \cite{1967SvA....10..602Z, Carr:1974nx}, ultra-compact mini-halos \cite{Ricotti:2009bs}, and other macroscopic objects, such as those of nuclear density \cite{Witten:1984rs, Lynn:1989xb}.

However, an important difference is that because $N$-MACHOs do not experience Hawking radiation \cite{Hawking:1971ei} they can be stable for arbitrarily small masses. This goes in contrast with PBHs with mass below $10^{-18}\,M_{\odot}$ whose abundance has been very strongly constrained due to Hawking evaporation. Thus, unlike microscopic black holes, light $N$-MACHOs can be stable in a much larger range of masses in which they could in principle constitute the entirety of the dark matter.

We must stress that although it is conceivable that either due to quantum-gravitational back reaction \cite{dvali2011black, Dvali_2016} or an assumed modifications of Hawking radiation \cite{Raidal:2018eoo} the stability window of black holes can be expanded, the $N$-MACHOs are special in the sense that their stability does not require going beyond standard semi-classical gravity.

\begin{figure}[t]
	\vs{1mm}
	\centering
	\includegraphics[scale= 1,angle=0]{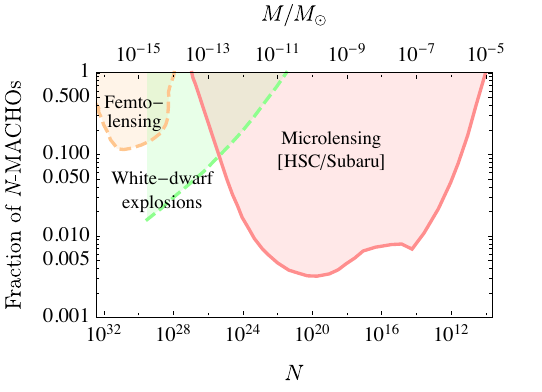}
	\caption{
		Constraints on the allowed $N$-MACHO DM fraction as a function of $N$.
		In the allowed range $10^{16} < N < 10^{32}$,
		there are bounds from femtolensing of gamma-ray bursts
		with the FERMI telescope \cite{2012PhRvD..86d3001B},
		from White-dwarf explosions \cite{2015PhRvD..92f3007G} and
		from the microlensing search of M31 stars
		based on HSC/Subaru data \cite{Niikura:2017zjd} (red contours; 95\% C.L.).
		It is assumed that the $N$-MACHO mass is given by $M_{\odot} / \sqrt{N\.}$
		[see Eq.~\eqref{eq:MofN}].\\[-7mm]
	}
	\label{fig:ExclusionN}
\end{figure}

In the following we will assume the previously mentioned $N$-neutron star to be the main representative of $N$-MACHOs. Furthermore we will again use the window for $N$ according to the proposed cosmological scenario in Ref.~\cite{Dvali:2009fw} modified by non-annihilating electrons, i.e., $10^{16} \leq N \leq~10^{32}$, so that the corresponding masses lie between $10^{-16}\,M_{\odot}$ and $10^{-8}\,M_{\odot}$. In this interval, there are lensing observations searching for compact objects:
\begin{itemize}
	\item[(\it i)]	femtolensing of gamma-ray bursts with the FERMI telescope \cite{2012PhRvD..86d3001B},
					posing constraints in the range $10^{-17}\,M_{\odot}$ and $10^{-14}\,M_{\odot}$, and
					
	\item[(\it ii)]	microlensing of M31 stars based on HSC/Subaru data \cite{Niikura:2017zjd},
					which limits the abundance of compact DM objects from
					$\sim 10^{-14}\,M_{\odot}$ to $\sim 10^{-5}\,M_{\odot}$.
\end{itemize}
Additionally, there are constraints coming from white-dwarf explosions. These are triggered by transit of compact bodies (such as PBHs) through the wide dwarfs, which then initiate thermonuclear fusion \cite{2015PhRvD..92f3007G}. This has been argued to rule out PBHs with masses $10^{-15}$ -- $10^{-11}\,M_{\odot}$. Undoubtably, this raises the intriguing possibility that a class of supernovae may be triggered in this way rather than via the conventional mechanism. However, it has recently been argued that this argument is inapplicable \cite{Montero-Camacho:2019jte}. Hence, we show these constraints in Fig.~\ref{fig:ExclusionN} with a broken line, reflecting the uncertainty on those limits. Similarly, the mentioned femtolensiong constraints recently got revised, where it has been argued that most gamma-ray bursts are inappropriate for femtolensing searches due to their large sizes \cite{Katz:2018zrn}. Therefore, we again display the corresponding limit with dashed line type.

\begin{figure}[th]
	\centering
	\includegraphics[scale= 1,angle=0]{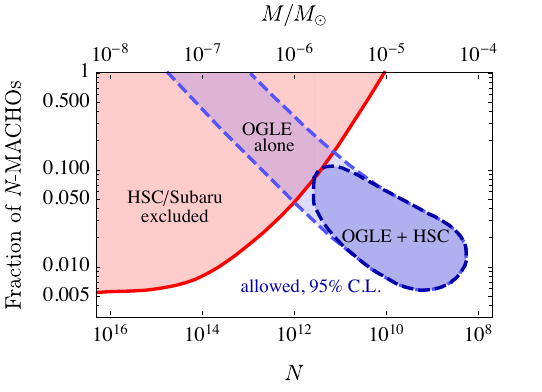}
	\caption{The blue-shading represents the 95\% C.L.~allowed region
		of $N$-MACHO abundance using combined data from OGLE \cite{2015AcA....65....1U} and
		HSC/Subaru \cite{Niikura:2017zjd}.
		(Figure adapted from Ref.~\cite{Niikura:2019kqi})\\[-6mm]
	}
	\label{fig:OGLE-Detection}
\end{figure}

In regard of the constraints coming from microlensing searches of M31 stars, a remark is in order: because of the black hole's significantly higher compactness, the respective Schwarzschild radius is (much) smaller than the wavelength of the lensed light used by the instruments. Therefore, Ref.~\cite{Niikura:2017zjd} actually poses weaker constraints on PBHs, so that there are essentially no constraints below masses of $\sim 10^{-12}\,M_{\odot}$. Despite this fact, even for $N$-MACHOs being (much) larger than black holes, there is an unconstrained mass gap in which these objects could, in principle, constitute all the DM. The allowed fraction of $N$-MACHOs is shown in Fig.~\ref{fig:ExclusionN} for the mass range $10^{-16} ~M_{\odot} < M < 10^{-7} ~M_{\odot}$. An interesting advantage of DM in this scenario is that it does not need to be entirely in the form of $N$-MACHOs, since they are composed of elementary dark particles.

As we stressed previously, the mentioned interval of allowed $N$-MACHO masses is derived under the premise that these are given by $M_{\odot} / \sqrt{N\.}$ . However, even if this was the case, there would still remain a possibility that not all sectors contribute (equally) to the $N$-MACHOs or that the sectors have different abundances, leading to heavier objects. This option is particularly appealing due to recent results using OGLE microlensing data \cite{Niikura:2019kqi}. Therein it has been found that PBHs actually provide the {\it best fit} (!) to the data explaining a population of six ultrashort-timescale events. Here, we would like to point out that these events could, of course, equally well be explained by $N$-MACHOs. We depict this in Fig.~\ref{fig:OGLE-Detection}.

As mentioned earlier, besides the lensing signatures discussed in this Section, $N$-MACHOs will lead to gravitational-wave emission, allowing to constrain their abundance. Similarly to PBHs, these objects would also emit gravitational waves below the mass window of astrophysical black holes which are constraint to $\gtrsim \Ocal ( 1 )\,M_{\odot}$. The gravitational waves from $N$-MACHO mergers should be different in two ways. Firstly, they suffer from a lack of energy due to radiation into particles from other sectors (see the discussion in Sec.~\ref{sec:Radiation-into-Dark-Photons}). Secondly, due to the $N$-MACHOs' vastly different mass-to-extend ratio, their waveforms and merger time will be significantly different, allowing to distinguish these objects from PBHs.

Also, similarly to PBHs, $N$-MACHOs will be subject to accretion around them. This concerns {\it (a)\.} SM particles from our sector, yielding the same accretion constraints from CMB distortions as PBHs (see Refs.~\cite{Ali-Haimoud:2016mbv, Nakama:2017xvq}); {\it (b)\.} in case of extensions to the SM, particles above the SM scale, implying enhanced annihilation signatures in the accreted dark matter spikes (\cf~Refs.~\cite{Eroshenko:2016yve, Boucenna:2017ghj}); {\it (c)\.} particles from other sectors, which would lead to effectively more extended, puffier objects with, for instance, different lensing characteristics. Of course, the above mentioned possibilities could occur in combination.

%%%%%%%%%%%%%%%%%%%%%%%%%%%%%%%%%%%%%%%%%%%%%%%%%%%%%%%%
%%%%%%%%%%%%%%%%%%%%%%%%%%%%%%%%%%%%%%%%%%%%%%%%%%%%%%%%
\section{Formation}
\label{sec:Formation}

%%%%%%%%%%%%%%%%%%%%%%%%%%%%%%%%%%%%%%%%%%%%%%%%%%%%%%%%
\subsection{Present-Day Formation}
\label{sec:Present-Day-Formation}

In general it is assumed that DM cannot possess a mechanism to dissipate energy, because this would violate the known shape of DM halos at scales of dwarf galaxies. Therefore, it may seem surprising that in the many-species model the entirety of DM could dissipate. However, given a large enough number of populated sectors, each sector becomes so dilute that the probability for the encounter of particles from the same species becomes extremely small. For example, in the extreme case of equally-populated $N = 10^{32}$ dark SM copies, today's mean partial density in any other sector would be $n_{j}^{0} \sim 10^{-32}\,n_\brm^{0} \sim 10^{-38}\cm^{-3}$, where $n_\brm$ denotes the baryonic density. This number is so dilute that the particles from any given dark SM copy have practically no chance of meeting each other. Also, notice that the energy loss by hidden baryons due to emission of dark photons at the galactic scale is negligible. Indeed, the energy loss due to a dark electromagnetic radiation by a hidden proton freely falling in a gravitational potential of the galaxy is,
	\begin{equation} \label{eq:rate}
 		\frac{ \d E }{ \d t }
 			\sim
 				\alpha\,\frac{ R_{\Grm}^{2} }{ R^{4} }
 			\sim
 				10^{-54}\,{ \erm \Vrm \over \sec }
 			\; ,
	\end{equation}
where $\alpha$ is electromagnetic coupling constant, ${R_{\Grm} \sim 10^{17}\cm}$ is the gravitational radius of our galaxy and $R \sim 50$\,kpc is its actual radius. Obviously, in such a case, the system would imitate collisionless dark matter.

Thus, the interesting regimes for $N$-MACHO formation are for larger over-densities or lower numbers of $N$ with higher population.

In the latter scenario, since the other sectors are composed of SM copies, one possibility is that the formation of $N$-MACHOs happens similar to usual star formation -- but time delayed. This was first pointed out in the context of folded braneworlds \cite{ArkaniHamed:1999zg} and later in the context of many SM copies \cite{Dvali:2009ne}. In order to understand this, we consider the time $t_{\rm cool}$ it takes a plasma to lose an $\Ocal( 1 )$ fraction of its kinetic energy. For a single sector this cooling time scales as $t_{\rm cool} \sim n^{-1}$. Since galaxy formation in our sector starts at a redshift $z \sim 4$, the galaxy formation in the dark sector would start right now if
	\begin{equation}
		\frac{ n_{j} }{ n_\brm }
			\lesssim
				\frac{ 1 }{ 4 }
			\; .
	\end{equation}
This scenario would require few dark sectors to be extremely more populated than the others in order to keep $\rho_{\rm DM}$ at the observed value. Then, according to App.~\ref{sec:N--MACHO-Solution-for-Unequal-Sectors},
	\begin{equation}
		\rho_{j}
			\ll
				\rho_{\rm max}
			\lesssim
				\rho_{\rm tot}
			\; ,
	\end{equation}
leading to an $N_{\rm eff}$ of $\Ocal( 1 )$. The resulting DM objects would hence have masses around $M_{\odot}$. This possibility, in particular, is very interesting, since it would lead to an increase in the number of gravitational-wave sources as a result of the violent collapse characteristic of star formation. However, it does not enable the formation of low mass $N$-MACHOs.

%%%%%%%%%%%%%%%%%%%%%%%%%%%%%%%%%%%%%%%%%%%%%%%%%%%%%%%%
\subsection{Primordial Formation}
\label{sec:Primordial--Formation}

As discussed above, in the present-day scenario, the $N$-MACHOs cool because the DM simply takes longer to dissipate enough energy. It is important to realise that the cooling mechanisms only operate efficiently within a certain window of masses set by the parameters of the theory. This is why in our sector baryons collapse and fragment below the characteristic mass scale $M_{\rm cr} \sim 10^{12}\,M_{\odot}$, while galaxy clusters are roughly spherical collections of virialized baryonic gas. A cooling mechanism is efficient if it is able to remove an $\Ocal( 1 )$ fraction of the kinetic energy within the characteristic free-fall time. Therefore, the baryonic cooling mechanism is not efficient on scales of clusters.

The same reasoning also applies for dissipative DM, possibly resulting in a much smaller value of $M_{\rm cr}$ \cite{Buckley:2017ttd}. In our scenario the only parameter that controls the efficiency of the cooling mechanism is the tiny partial density of the dark sectors, which in turn depends on the number of populated copies $N$. Thus, for regions where a high concentration of individual sectors is achieved, the cooling can become efficient.

With that in mind we propose that the necessary seed overdensity for the formation of $N$-MACHOs could also be primordial. This possibility shares similarities with the formation of so-called ultra-compact mini-halos \cite{Ricotti:2009bs}. These are compact structures, whose origin are primordial inflationary overdensities reentering the horizon during radiation domination. The overdensities need to be relatively large but below the threshold for primordial black-hole formation \cite{1967SvA....10..602Z, Carr:1974nx}, meaning that the necessary overdensities $\delta \equiv \delta \rho / \rho$ lie in the range $\delta_{\rm min} \sim 10^{-4} < \delta < \delta_{\rm max} \sim 10^{-1}$. Here, the lower threshold depends on the scale and on the value of the critical redshift $z_{\rm cr}$ until these structures are essentially unaltered and undergo gravitational collapse. Very roughly, this holds true until a non-linear structure formation starts. The conservative scenarios assume $z_{\rm cr} = 1000$, but it may as well be an order of magnitude lower. In such a case, their fraction can be many orders of magnitude higher at the present day. It can be shown \cite{Ricotti:2009bs} that during radiation domination those overdensities grow logarithmically, which later during the matter domination changes to a linear growth till the collapse starts.

While in the standard DM scenario the ultra-compact mini-halos form virialized structures, the situation changes fundamentally in the large $N$ species framework. Due to an efficient dissipation of energy on those small scales, the ultra-compact mini-halos could cool and serve as formation sites of $N$-MACHOs. Their (primordial) fraction can be estimated assuming a Gaussian distribution via
\begin{align} \label{eq:betaOfEll}
	\beta( \ell )
		&=
			\frac{ 1 }{ \sqrt{2\pi\.}\.\sigma_{\Hrm}( \ell )}\,
			\int_{\delta_{\rm min}}^{\delta_{\rm max}}
			\d \delta\;
			\exp\!
			\left[
				-
				\frac{ \delta^{2} }{ 2\.\sigma_{\Hrm}^{2}( \ell ) }
			\right]
		,
	\end{align}
which is the probability that a region of co-moving size $\ell$ (at the time when $1 / \ell = a\mspace{1mu}H$, i.e., when this scale reenters the Hubble horizon) will later undergo a collapse to an $N$-MACHO. Above, $\sigma_{\Hrm}^{2}$ is the DM variance at the time of horizon entry. The quantity $\beta$ is exponentially sensitive to the value of the lower threshold $\delta_{\rm min}$. Unless this becomes of the order $10^{-6}$ -- which it does for $z_{\rm cr} \sim \Ocal( 1 )$ -- the standard almost scale-invariant expression for the primordial power spectrum with moderate tilt, yields only a negligible fraction of primordial-origin $N$-MACHOs.

In order to make this fraction non-negligible, a mechanism for increasing the primordial power spectrum has to be employed. Notice, this is the same condition that is required for PBH formation (see Ref.~\cite{Carr:2016drx} for a review). In this sense, formation conditions for $N$-MACHOs are not any less natural than the ones demanded by PBH. We want to stress, however, that the power spectrum required for obtaining a substantial fraction of $N$-MACHOs is much easier to achieve than in the PBH case, due to a much lower threshold.

An interesting alternative for increasing the $N$-MACHO fraction is given by extending the matter-dominated era due to reheating between inflation and radiation domination. In this scenario as much as 50\% of all DM could be in compact DM structures at $z = 100$ \cite{Erickcek:2011us}. Coincidentally, since the reheating temperature is proportional to $N^{-1/2}$ according to Ref.~\cite{Dvali:2009fw}, the values at the larger end of the range of $N$ automatically lead to such a phase.

Here, a cautionary note is necessary. In the cosmological scenario in Ref.~\cite{Dvali:2009fw}, the dark particles never reach thermal equilibrium due to their tiny densities. The advantage of this is that the predictions in the dark sectors are independent of the concrete mechanism of baryon asymmetry generation in our sector.
 The other copies, however, are composed of effectively an equal number of baryons and anti-baryons. Within that scenario, the right DM abundance is produced provided that all annihilation channels are always frozen out. Regions of higher density could not only spoil this prediction, but also strongly suppress the fraction of $N$-MACHOs as a consequence of favoring an annihilation over a dissipation.

%%%%%%%%%%%%%%%%%%%%%%%%%%%%%%%%%%%%%%%%%%%%%%%%%%%%%%%%
%%%%%%%%%%%%%%%%%%%%%%%%%%%%%%%%%%%%%%%%%%%%%%%%%%%%%%%%
\section{Conclusion \& Outlook}
\label{sec:Conclusion-and-Outlook}

The purpose of the present paper was to point out some astrophysical consequences of scenarios in which DM originates from species of a large number $N$ of hidden sectors. At the fundamental level, such theories are motivated by the solution of the Hierarchy Problem \cite{Dvali:2007hz, Dvali:2007wp} but as a bonus 
 they offer potentially interesting DM scenarios \cite{Dvali:2009ne, Dvali:2009fw}. In this picture, the DM is composed out of particles belonging to many dark sectors. Each sector is so dilute that particles belonging to it rarely meet each other. This creates an effect of slow dissipation, despite the fact that at the fundamental level the DM species can posses interactions that are as strong as the ones experiences by the Standard Model particles. This interplay between low partial densities and unsuppressed forces results into unusual composite structures that can be of both fundamental as well as astrophysical significance.

Using the Tolman-Oppenheimer-Volkoff equation, we showed that particles from different dark sectors can collectively form stable compact objects, which we called $N$-MACHOs. As a direct consequence of the segmentation into $N$ mutually non-interacting sectors, these objects have very low mass and a tiny radius as compared to the single-sector case. This can be intuitively understood by the reduction of the thermal pressure counteracting the gravitational collapse, in contrast to the unchanged gravitational force. We showed under general assumptions that this holds for different abundances and masses in the other sectors.

For the scenario where the dark matter is represented by $N$ hidden copies of the SM, these solutions do not rely on a new mechanism, but rather use the same well-understood properties of known stellar structure. There, we argue that $N$-MACHOs could form in high-density regions as a consequence of an additional feature in the inflationary spectrum or a long matter-dominated phase between inflation and radiation domination. The latter possibility is of particular interest, since such a phase automatically appears for large values of $N$ in the underlying cosmological scenario \cite{Dvali:2009fw}.

The case with $N$ identical copies of the SM is particularly appealing due to the fact that structures composed of SM particles are relatively well understood. It would be interesting, however, to embed $N$-MACHOs in other scenarios. Since the only real requirement for $N$-MACHOs is a polytropic equation of state, a bosonic realization could be constructed, where the particles would form a weakly interacting Bose gas. Moreover, for particles of higher mass, $N$-MACHOs could have masses below $10^{-16}\,M_{\odot}$. Since they are not subject to Hawking evaporation, a totally unconstrained region could hence be entered, where such objects could make up all of the dark matter.
%This could be of particular interest for supersymmetric incarnations of the large-$N$ species model.

%%%%%%%%%%%%%%%%%%%%%%%%%%%%%%%%%%%%%%%%%%%%%%%%%%%%%%%%
%%%%%%%%%%%%%%%%%%%%%%%%%%%%%%%%%%%%%%%%%%%%%%%%%%%%%%%%
\appendix

%%%%%%%%%%%%%%%%%%%%%%%%%%%%%%%%%%%%%%%%%%%%%%%%%%%%%%%%
%%%%%%%%%%%%%%%%%%%%%%%%%%%%%%%%%%%%%%%%%%%%%%%%%%%%%%%%
\section{Newtonian Polytropes}
\label{sec:Newtonian-Polytropes}

For a polytrope, the TOV equation can be brought into a dimensionless form by introducing new variables $\theta$ and $\xi$, defined by
	\begin{subequations}
	\begin{align}
		r
			&=
				\!\left(
					\frac{ 2 M_{\Prm}^{2}\.K_{\rm tot}\.\gamma }{ \gamma - 1 }
				\right)^{\mspace{-5mu}1 / 2}
				\rho_{\rm tot}( 0 )^{(\gamma - 2)/2}\;\xi
			\; ,
			\label{eq:rad}
			\displaybreak[1]
			\\[1.5mm]
		\rho_{\rm tot}( r )
			&=
				\rho_{\rm tot}( 0 )\;\theta^{1 / ( \gamma - 1 ) }( r )
			\; ,
			\label{dens}
			\displaybreak[1]
			\\[1.5mm]
		p_{\rm tot}( r )
			&=
				K_{\rm tot}\.\rho_{\rm tot}( 0 )^{\gamma}\;
				\theta^{\gamma / ( \gamma - 1 )} ( r )
			\; ,
	\end{align}
	\end{subequations}
yielding
	\begin{equation}
		\frac{ 1 }{ \xi^{2} }
		\frac{ \d }{ \d\xi }\!
		\left(
			\xi^{2}
			\frac{ \d\theta }{ \d\xi }
		\right) +
		\theta^{1 / ( \gamma - 1 )}
			=
				0
			\; ,
			\quad\;
		\theta( 0 )
			=
				\theta'( 0 )
			=
				0
			\; .
	\end{equation}	
The function $\theta( \xi )$ defined by this equation is called the Lane-Emden-function of index $( \gamma - 1 )^{-1}$. It can be shown \cite{Weinberg1972-WEIGAC}, that for $\gamma > 6 / 5$, $\theta( \xi )$ vanishes at some finite $\xi_{0}$. This fact and Eq.~\eqref{eq:rad} can be used to define the radius of the object by
	\begin{equation} \label{eq:PolR}
		R
			\equiv
				\!\left(
					\frac{ 2\.M_{\Prm}^{2}\.K_{\rm tot} \gamma }{ \gamma - 1 }
				\right)^{\mspace{-5mu}1 / 2}
				\rho_{\rm tot}( 0 )^{( \gamma - 2 ) / 2}\.\xi_{0}
			\; .
	\end{equation}
Using this radius, Eqs.~(\ref{eq:rad},b), the mass of the polytrope can be calculated to be
	\begin{align}\nonumber
		M
			&\equiv
				\int_{0}^{R}\d r\;
				4\pi\.r^{2}\.\rho_{\rm tot}( r )
			\displaybreak[1]
				\\[1mm]
			&=
				4\pi\.\rho_{\rm tot}( 0 )^{ ( 3 \gamma - 4 ) / 2 }
				\left(
					\frac{ 2 M_{\Prm}^{2}\.K_{\rm tot}\.\gamma }{ \gamma - 1 }
				\right)^{\mspace{-5mu}3 / 2}
				\xi_{0}^{2}\,
				\big|
					\theta'(\xi_{0})
				\big|
				\; .
	\label{eq:PolM}
	\end{align}
The numerical values for $\xi_{0}$ and $-\,\xi_{0}^{2}\.\theta'( \xi_{0} )$ for the case of $\gamma = 5 / 3$ are given by $\xi_{0} = 3.65375$ and $-\,\xi_{0}^{2}\.\theta'( \xi_{0} ) = 2.71406$ (see for instance Ref.~\cite{Weinberg1972-WEIGAC} for further details and more numerical values).

%%%%%%%%%%%%%%%%%%%%%%%%%%%%%%%%%%%%%%%%%%%%%%%%%%%%%%%%
%%%%%%%%%%%%%%%%%%%%%%%%%%%%%%%%%%%%%%%%%%%%%%%%%%%%%%%%
\acknowledgments

This work was supported in part by the Humboldt Foundation under Humboldt Professorship Award, by the Deutsche Forschungsgemeinschaft (DFG, Ger- man Research Foundation) under Germany's Excellence Strategy - EXC-2111 - 390814868, and Germany's Excellence Strategy under Excellence Cluster Origins. F.K.~acknowledges support from the Swedish Research Council (Vetenskapsr{\r{a}}det) through contract No.~638-2013-8993 and the Oskar Klein Centre for Cosmoparticle Physics. He also thanks the Arnold Sommerfeld Center at Ludwig-Maximilians-Universit{\"a}t for hospitality received during part of this work.

%%%%%%%%%%%%%%%%%%%%%%%%%%%%%%%%%%%%%%%%%%%%%%%%%%%%%%%%%%%
%%%%%%%%%%%%%%%%%%%%%%%%%%%%%%%%%%%%%%%%%%%%%%%%%%%%%%%%
\setlength{\bibsep}{5pt}
\setstretch{1}
\bibliographystyle{utphys}
\bibliography{refs}

\end{document}